\documentclass{article}

\usepackage{amsmath,amssymb,stmaryrd,fullpage,times}
\usepackage{tikz,url,cite}

\newcommand{\ket}[1]{\left|#1\right>}

\newtheorem{remark}{{\bf Remark}}[section]
\newtheorem{proposition}{{\bf Proposition}}[section]
\newtheorem{corollary}{{\bf Corollary}}[section]
\newtheorem{example}{{\bf Example}}[section]
\newtheorem{definition}{{\bf Definition}}[section]

\makeatletter
\def\mop#1{\mathop{\operator@font {#1\null}}}
\makeatother
\def\bigO{{\mop{O}}}
\def\HW{{\mop{HW}}}
\def\ord{{\mop{ord}}}

\begin{document}
\title{Quantum binary field inversion: improved circuit\\ depth 
via choice of basis representation}

\author{Brittanney Amento\\
Florida Atlantic University\\
Department of Mathematical Sciences\\
Boca Raton, FL 33431\\
{\tt bferoz@fau.edu}\and
Martin R{\"o}tteler\\
NEC Laboratories America\\
4 Independence Way, Suite 200\\
Princeton, NJ 08540, U.S.A.\\
{\tt mroetteler@nec-labs.com}
\and Rainer Steinwandt\\
Florida Atlantic University\\
Department of Mathematical Sciences\\
Boca Raton, FL 33431\\
{\tt rsteinwa@fau.edu}
}

\maketitle

\begin{abstract}
Finite fields of the form ${\mathbb F}_{2^m}$ play an important role in coding theory and cryptography. We show that the choice of how to represent the elements of these 
fields can have a significant impact on the resource requirements for quantum 
arithmetic. In particular, we show how the use of Gaussian normal basis representations and of `ghost-bit basis' representations can be used to implement inverters with a quantum circuit of depth $\bigO(m\log(m))$. To the best of our knowledge, 
this is the first construction with subquadratic depth reported in the literature. 
Our quantum circuit for the computation of multiplicative inverses is based on the Itoh-Tsujii algorithm which exploits that in normal basis representation squaring corresponds to a permutation of the coefficients. We give resource estimates for the resulting quantum circuit for inversion over binary fields ${\mathbb F}_{2^m}$ based on an elementary gate set that is useful for fault-tolerant implementation.
\end{abstract}

\section{Introduction}

In quantum computing, arithmetic operations occur in a plurality of contexts \cite{Sho97,Kitaev:97,Hallgren:2002,DCW:2002,vDHI:2003,CSV:2007,MRR+:2007}. Having good quantum circuits for arithmetic is indispensable for obtaining good resource estimates and efficient circuit implementations of more complex quantum algorithms. In view of the cryptographic significance, it is not surprising that a number of publications have already explored quantum circuits to implement finite field arithmetic, including \cite{BBF03,KaZa04,MMCP09,MMCP09b}. Important special cases are arithmetic operations in finite prime fields and finite binary fields (cf., for instance, \cite{FIPS1863}). While there is some common ground between the prime-field case and the characteristic-two case, there are also important differences. In this paper we focus entirely on quantum circuits to implement arithmetic in fields of the form ${\mathbb F}_{2^m}$.

Interestingly, thus far the literature on quantum circuits for ${\mathbb F}_{2^m}$-arithmetic focuses completely on polynomial basis representations, and computing multiplicative inverses by implementing the extended Euclidean algorithm as discussed in \cite{KaZa04} appears to be the common choice. The cost of implementing inversion this way is significant as the resulting circuit has a size that is cubic in $m$. When realizing the group law on a binary elliptic curve as quantum circuit, the cost of this operation becomes apparent: in an earlier issue of this journal, Maslov et al. presented a solution to the discrete logarithm problem on binary elliptic curves \cite{MMCP09b}. An important technique for achieving quadratic depth with their solution was to bring down the number of finite field inversions to one. For the asymptotic analysis, the quadratic depth of this single inversion is still as expensive as all other arithmetic operations combined. So when trying to improve on the discrete logarithm circuit presented in \cite{MMCP09b}---which from a cryptanalytic point of view is desirable---reducing the complexity of binary finite field inversion is a natural first step. 

\paragraph{Our contribution.} This paper presents linear-depth multipliers using a so-called ghost-bit basis and using Gaussian normal bases. Building on these multipliers, we describe an inverter for ${\mathbb F}_{2^m}^*$ of depth $\bigO(m\log(m))$ derived from a classical inversion algorithm by Itoh and Tsujii \cite{ItTs89}, using $\bigO(m\log(m))$ qubits. We hope that our work stimulates follow-up work on using different representations of finite fields in quantum circuits, and we expect that the circuits presented in this paper will be useful for speeding 
up the arithmetic for quantum algorithms for computing discrete logarithms on
elliptic curves, but also for other algebraic problems that can be tackled 
on a quantum computer, including hidden polynomial equations \cite{CSV:2007}, hidden shift problems \cite{vDHI:2003,Roetteler:2010,ORR:2012}, and certain period finding tasks \cite{Sho97,Kitaev:97,Hallgren:2002}. 

For the fault-tolerant implementation of quantum circuits on several error-correcting 
codes \cite{ASG:2009,Reichardt:2009} the elementary gate set consisting of all Clifford gates
and the so-called $T$-gate is a preferable one. The $T$-gate is the local unitary ${\rm diag}(1,\exp(2\pi i/8))$. The actual complexity of a fault-tolerant implementation of $T$-gates is extremely high, hence it is preferable to reduce their number as much as possible. We show that in a Gaussian normal basis or a ghost-bit basis representation, an inversion over ${\mathbb F}_{2^m}$ can be computed in a $T$-depth of $\bigO(m \log(m))$ and using at most $\bigO(m^2 \log(m))$ many $T$-gates.

\section{Preliminaries: finite fields ${\mathbb F}_{2^m}$}
Perhaps the most popular representation of finite fields ${\mathbb F}_{2^m}$ is the use of a polynomial basis. In the following, we briefly review some basic facts about this representation as well as two alternatives---the use of a ghost-bit basis and of a Gaussian normal basis. All of these representations are known, and we claim no originality for this section. 

\subsection{Polynomial basis representation}
 Denoting by $f=x^m+\sum_{i=0}^{m-1}x^i\in{\mathbb F}_2[x]$ an irreducible polynomial of degree $m$ over the prime field ${\mathbb F}_2$, we can identify ${\mathbb F}_{2^m}$ with the quotient ring ${\mathbb F}_2[x]/(f)$, and this identification forms the basis of a popular representation of binary finite fields.

\begin{definition}[Polynomial basis representation]\label{def:polybasis}\ \\
With the above notation, let $x^0+(f),x^1+(f),\dots,x^{n-1}+(f)$ be the canonical ${\mathbb F}_2$-vector space basis of ${\mathbb F}_2[x]/(f)$. In the \emph{polynomial basis representation}, each $\alpha\in{\mathbb F}_{2^m}$ is represented by the unique tuple $(\alpha_0,\dots,\alpha_{m-1})\in{\mathbb F}_{2}^m$ such that $\alpha=\sum_{i=0}^{m-1}\alpha_i\cdot (x^i+(f))$.
\end{definition}

\begin{example}\label{exa:exampleone}
The polynomial $x^4+x^3+x^2+x+1\in{\mathbb F}_2[x]$ is irreducible, and so the field with $16$ elements can be identified with ${\mathbb F}_2[x]/(x^4+x^3+x^2+x+1)$. Choosing $f=x^4+x^3+x^2+x+1$ in the above definition, in the polynomial basis representation, the tuple $(1,0,1,0)\in{\mathbb F}_2^4$ represents the field element $x^2+1+(f)$.
\end{example}
In the current literature on quantum arithmetic for binary finite fields, the representation from Definition~\ref{def:polybasis} seems to be the only one considered. Beauregard et al. \cite{BBF03}, Maslov et al. \cite{MMCP09}, and Kaye and Zalka \cite{KaZa04} provide circuits for addition, multiplication and inversion using a polynomial basis.
\begin{itemize}
   \item Using one qubit per coefficient  of $\alpha=\sum_{i=0}^{m-1}\alpha_i\cdot (x^i+(f))$, adding $\ket{\alpha}$ to an $m$-qubit input $\ket{\beta}$ can be done in the obvious way with $m$ CNOT gates, each conditioned on one of the $\alpha_i$. These CNOT gates operate on disjoint wires, and hence this adder can be realized in depth $1$.
  \item Building on a classical Mastrovito multiplier \cite{Mas88,Mas91,MaHa04}, the multiplication of two $m$-qubit inputs $\ket{\alpha}$ and $\ket{\beta}$ can be realized in depth $9m+\bigO(1)$ using Toffoli gates.
 If the irreducible polynomial $f$ is the all-one polynomial or a trinomial, $m^2-m-1$ gates suffice \cite{MMCP09}.
  \item Computing the inverse of a non-zero $\alpha\in{\mathbb F}_{2^m}$, using the extended Euclidean algorithm, can be implemented in depth $\bigO(m^2)$ and $2m+\bigO(\log(m))$ qubits \cite{KaZa04,MMCP09b}.
\end{itemize}
In this paper, we will look at two different representations of binary fields which---from an algorithmic point of view---suggest an interesting alternative to the use of a polynomial basis.

\subsection{Ghost-bit basis representation}
Keeping the notation from above, suppose the irreducible polynomial $f$ we use is the all-one polynomial $x^m+\dots+1$. In this case, $m+1$ is prime and $2$ is a generator of the cyclic group ${\mathbb F}_{2^{m+1}}^*$ (cf. \cite{ItTs89}). Then $f$ divides $x^{m+1}+1=(x+1)\cdot(x^m+\dots+1)\in{\mathbb F}_2[x]$, and we can define the map
\begin{equation*}
\begin{array}{lccc}
\phi:&{\mathbb F}_{2}[x]/(f)&\longrightarrow&{\mathbb F}_{2}[x]/(x^{m+1}+1)\\
&\sum_{i=0}^{m-1}\alpha_i\cdot x^i+(f)&\longmapsto&\sum_{i=0}^{m-1}\alpha_i\cdot x^i+(x^{m+1}+1)
\end{array}.
\end{equation*}
The map $\phi$ may be seen as appending an extra (zero) bit to the coefficient vector of a polynomial basis representation of $\alpha\in{\mathbb F}_{2}[x]/(f)$. As detailed by Silverman \cite{Sil99} (who suggests to attribute the construction to Itoh and Tsujii \cite{ItTs89}), instead of adding, multiplying, and inverting elements in ${\mathbb F}_{2}[x]/(f)$ directly, we can apply $\phi$ to the operands, perform the needed additions, multiplications, and inversions in ${\mathbb F}_{2}[x]/(x^{m+1}+1)$, and then map the result back into ${\mathbb F}_{2}[x]/(f)$ by applying
\begin{equation}\label{equ:backfromgbb}
\begin{array}{lccc}
&{\mathbb F}_{2}[x]/(f)&\longleftarrow&{\mathbb F}_{2}[x]/(x^{m+1}+1)\\
&\sum_{i=0}^{m-1}(\alpha_i+\alpha_m)\cdot x^i+(f)&\longmapsfrom&\sum_{i=0}^{m}\alpha_i\cdot x^i+(x^{m+1}+1)
\end{array}.
\end{equation}

\begin{definition}[Ghost-bit basis representation]\ \\
With the above notation, assume that $1+\dots+x^m$ is irreducible.
In the \emph{ghost-bit basis representation}, each $\alpha$ is represented by a tuple $(\alpha_0,\dots,\alpha_{m})\in{\mathbb F}_2^{m+1}$ such that $(\alpha_0+\alpha_m,\dots,\alpha_{m-1}+\alpha_m)$ is the polynomial basis representation of $\alpha$ using the irreducible polynomial $1+\dots+x^m$.
\end{definition}

Thence, a conversion from the \emph{ghost-bit basis representation} to a polynomial basis representation boils down to dropping the ghost bit and adding (XOR) it to the remaining $m$ bits. In a quantum circuit, this translates into a single CNOT with multiple fan-out at the very end, provided we do not have to restore the initial $\ket{0}$-value of the ghost (qu)bit.
 We note that for adding field elements alone, applying the map $\phi$ has no advantage---but also no dramatic drawback.
\begin{itemize}
   \item Using one qubit per coefficient  of $\alpha=\sum_{i=0}^{m}\alpha_i\cdot x^i+(x^{m+1}+1)$, adding $\ket{\alpha}$ to an $(m+1)$-qubit input $\ket{\beta}$ can be done in the obvious way with $m+1$ CNOT gates, conditioned on the individual $\alpha_i$. These CNOT gates operate on disjoint wires, and hence this adder can be realized in depth $1$.
\end{itemize}
To realize quantum circuits for multiplying and inverting field elements, we are interested in exploiting the following properties of ${\mathbb F}_{2}[x]/(x^{m+1}+1)$:
\begin{itemize}
   \item Squaring corresponds to a shuffle of the coefficient vector:
         \begin{equation}\left(\sum_{i=0}^{m}\alpha_i\cdot x^i+(x^{m+1}+1)\right)^2=\sum_{i=0}^{m}\alpha_{\pi^{-1}(i)}\cdot x^i+(x^{m+1}+1),\label{equ:gbbsquare}
         \end{equation}
         where $\pi(i)=2\cdot i\bmod (m+1)$ for $i=0,\dots,m$.
\begin{example}\label{exa:gbb}
    As noted in Example~\ref{exa:exampleone}, the polynomial $x^4+x^3+x^2+x+1\in{\mathbb F}_2[x]$ is irreducible, and so ${\mathbb F}_{2^4}$ affords a ghost-bit basis representation: the above map $\phi$ translates operations in ${\mathbb F}_{2^4}$ into operations in ${\mathbb F}_2[x]/(x^5+1)$. Applying $\phi$ to $x^2+1+(x^4+x^3+x^2+x+1)$, we obtain $x^2+1+(x^5+1)$, i.\,e., the polynomial basis representation $(1,0,1,0)$ from Example~\ref{exa:exampleone} translates into the ghost-bit basis representation $(1,0,1,0,0)$.

For $m=4$, the permutation $\pi$ in Equation~\eqref{equ:gbbsquare} is $(0)(1,2,4,3)$, so the ghost-bit basis representation of $(x^2+1+(x^5+1))^2$ is $(1,0,0,0,1)$---corresponding to $x^4+1+(x^5+1)$. Applying the map from Equation~\eqref{equ:backfromgbb}, we obtain the corresponding polynomial basis representation $(1,1,1,0)$ respectively $x^3+x^2+x+(x^4+x^3+x^2+x+1)$.
\end{example}

   \item To multiply two elements $\alpha=\sum_{i=0}^{m}\alpha_i\cdot x^i+(x^{m+1}+1)$ and $\beta=\sum_{i=0}^{m}\beta_i\cdot x^i+(x^{m+1}+1)$, the following formula for the coefficients of their product $\gamma=\sum_{i=0}^{m}\gamma_i\cdot x^i+(x^{m+1}+1)$ can be used:
\begin{equation}
  \gamma_i=\sum_{j=0}^m\alpha_j\beta_{(i-j)\bmod{(m+1)}}\label{equ:gbbformula}
\end{equation}
\end{itemize}
As explained in Section~\ref{sec:MastrovitoGhostbit} below, in combination with an observation in \cite{MMCP09}, Equation~\eqref{equ:gbbformula} yields a linear-depth circuit for multiplication in ${\mathbb F}_2[x]/(x^{m+1}+1)$.

\begin{remark}The idea of a ghost-bit basis can be generalized to a representation with more redundancy---whenever the polynomial $x^n+1\in{\mathbb F}_2[x]$ has an irreducible factor $f$ of degree $m$, then we can define a map $\phi$ analogously as above, using $n-m$ `ghost bits.' Geiselmann and Lukhaub  \cite{GeLu03} discuss the implementation of ${\mathbb F}_{2^m}$-multiplication in such a representation with a classical reversible circuit.
\end{remark}

\subsection{Normal basis representation}

The possibility of an inexpensive squaring operation will be of great benefit for the inversion algorithm below, and a natural type of field representation to be considered in this context is a \emph{normal basis representation}.
\begin{definition}[Normal basis representation]\ \\
Let $\eta\in{\mathbb F}_{2^m}$ be such that $\{\eta,\eta^2,\eta^{2^2},\dots,\eta^{2^{m-1}}\}$ is an ${\mathbb F}_2$-vector space basis of ${\mathbb F}_{2^m}$.
In a \emph{normal basis representation} of ${\mathbb F}_{2^m}$, we represent each $\alpha\in{\mathbb F}_{2^m}$ by the unique tuple $(\alpha_0, \alpha_1, \cdots, \alpha_{m-1})\in{\mathbb F}_2^m$ with $\alpha=\sum_{i=0}^{m-1}\alpha_i\cdot(\eta^{2^i})$.
\end{definition}
 A normal basis representation exists for every field ${\mathbb F}_{2^m}$ of degree $m\ge 1$, and more background information on normal bases can be found in \cite{Jun93}, for instance. By construction, squaring in such a representation is just a cyclic shift, and addition can be implemented as bit-wise addition---just as in the case of a polynomial or ghost-bit basis representation. To ensure the availability of an efficient multiplication procedure, one often restricts to a particular type of normal basis, which exists whenever $8\nmid m$. In this paper we focus entirely on these so-called \emph{Gaussian normal bases}; see also \cite{JMV01,DHH06} for 
further background and proofs of the properties that are relevant for our purposes.

\begin{definition}[Gaussian normal basis]\label{def:GNB}\ \\
Assume that $t\ge 1$ such that $p = tm + 1$ is prime and the index of the subgroup generated by $2\in{\mathbb F}_p^*$ is coprime to $m$. Let $\alpha\in{\mathbb F}_{2^{mt}}$ be a primitive $p$-th root of unity, and let $u\in{\mathbb F}_p^*$ have order $t$.
 Then $$\left\{\sum_{j=0}^{t-1}\alpha^{u^j},\left(\sum_{j=0}^{t-1}\alpha^{u^j}\right)^{2^1},\dots,\left(\sum_{j=0}^{t-1}\alpha^{u^j}\right)^{2^{m-1}}\right\}$$ is a normal basis of ${\mathbb F}_{2^m}$, commonly referred to as \emph{type $t$ Gaussian normal basis}.\footnote{The basis elements are known as \emph{Gauss periods of type $(m,t)$}, but we do not need this terminology here.}
\end{definition}
The complexity of multiplication with respect to a Gaussian normal basis representation is reflected by its type $t$. The Digital Signature Standard \cite[Appendix~D.1.3]{FIPS1863} offers several practical examples for (extension degree, type)-pairs of binary fields ${\mathbb F}_{2^m}$: $(163,4)$, $(233,2)$, $(283,6)$, $(409,4)$, and $(571,10)$. For cryptographic applications, one is interested in situations where the type $t$ is small. Hence, in our analysis we regard $t$ as a (small) constant.

\begin{itemize}
   \item Using one qubit per coefficient of $\alpha$, adding $\ket{\alpha}$ to an $m$-qubit input $\ket{\beta}$ can be done in the obvious way with $m$ CNOT gates, conditioned on the individual $\alpha_i$. These CNOT gates operate on disjoint wires, and hence this adder can be realized in depth $1$.

   \item Squaring corresponds to a cyclic (right-)shift of the coefficient vector:
$$\begin{array}{ccc}
{\mathbb F}_{2^m}&\longrightarrow&{\mathbb F}_{2^m}\\
\sum_{i=0}^{m-1}\alpha_i\eta^{2^i}&\longmapsto&\sum_{i=0}^{m-1}\alpha_{i-1 (\bmod m)}\eta^{2^i}
\end{array}$$

	\item With the notation from Definition~\ref{def:GNB}, define $F(1), F(2), \ldots, F(p-1)$ through $F(2^iu^j \bmod p)=i$ for $0\leq i<m$ and $0\leq j<t$. Then the representation $(\gamma_0,\dots,\gamma_{m-1})$ of the product $\gamma=\alpha\cdot\beta$ can be computed as $\gamma_i=$
\begin{equation}
\left\{\begin{array}{l@{\hspace*{0em}}l}
\sum\limits_{k=1}^{tm-1}\alpha_{F(k+1)+i}\beta_{F(p-k)+i}&\text{, if }2\mid t\\
\sum\limits_{k=1}^{tm-1}\alpha_{F(k+1)+i}\beta_{F(p-k)+i}+\sum\limits_{k=1}^{m/2}(\alpha_{k-1+i}\beta_{k-1+\frac{m}{2}+i}+\alpha_{k-1+\frac{m}{2}+i}\beta_{k-1+i})&\text{, if }2\nmid t
\end{array}\right.\label{equ:gnbmult}
\end{equation}
for $i=0,\dots,m-1$ (with all indices being understood modulo $m$).
\begin{example}[Gaussian normal basis]\label{exa:gnb}
   For ${\mathbb F}_{2^5}$ there exists a Gaussian normal basis of type $t=2$ : we have $p=2\cdot 5+1=11$, and $2$ is a generator of ${\mathbb F}_p^*$, so the index of the subgroup generated by $2\in {\mathbb F}_p^*$ is certainly coprime to $m=5$. Choosing $u=10\in{\mathbb F}_{11}^*$ as an element of order $t=2$, we compute
$$\begin{array}{c|c|c|c|c|c|c|c|c|c}
 F(1)&F(2)&F(3)&F(4)&F(5)&F(6)&F(7)&F(8)&F(9)&F(10)\\\hline 
   0 & 1  & 3  &  2 & 4  & 4  & 2  & 3  & 1  & 0
  \end{array}\quad.$$
Now, from Equation~(\ref{equ:gnbmult}) for the general multiplication $\gamma=\alpha\cdot\beta$, we obtain
\begin{eqnarray}\gamma_i&=&\alpha_{1+i}\beta_i + \alpha_{3+i}\beta_{1+i} +\alpha_{2+i}\beta_{3+i} +\alpha_{4+i}\beta_{2+i} +\alpha_{4+i}\beta_{4+i} +\nonumber\\&&\alpha_{2+i}\beta_{4+i} +\alpha_{3+i}\beta_{2+i} +\alpha_{1+i}\beta_{3+i} +\alpha_i\beta_{1+i}\label{exa:gnbbeta}\end{eqnarray}
for $i=0,\dots,m-1$.
\end{example}
\end{itemize}

\subsection{Computing multiplicative inverses with the Itoh-Tsujii algorithm}\label{sec:itsutsuji}
With a field representation where squaring is inexpensive, looking at an exponentiation-based alternative to Euclid's algorithm for computing multiplicative inverses becomes worthwhile. For any $\alpha\in{\mathbb F}_{2^m}^*$, we have $\alpha^{2^m-1}=1$, and hence $\alpha^{-1}=\alpha^{2^m-2}$ can be found by raising $\alpha$ to the power $2^m-2$. The almost maximal Hamming weight of the latter makes a naive square-and-multiply implementation problematic. Happily, a technique by Itoh and Tsujii \cite{ItTs89} enables an efficient implementation of this exponentiation (see, e.\,g., \cite{ItTs89,TaTa01,RHSCC05,Gua11}). We begin by writing $$m-1=\sum_{i=1}^{\HW(m-1)}2^{k_i}\quad\text{, where }\lfloor\log_2(m-1)\rfloor=k_1>k_2>\cdots>k_{\HW(m-1)}\ge 0,$$ and $\HW(\cdot)$ denotes the Hamming weight. Now, for fixed $\alpha\in{\mathbb F}_{2^m}^*$ and for $i\ge 0$, we define $\beta_i=\alpha^{2^i-1}$. In particular, $\beta_0=1$, $\beta_1=\alpha$, and the inverse of $\alpha$ can be obtained as $\alpha^{-1}=(\beta_{{m-1}})^2$.
So once we know $\beta_{m-1}$, only one final squaring is needed---which for a ghost-bit or a normal basis representation is just a permutation. To compute $\beta_{m-1}$, we exploit the fact that for all non-negative integers $i,j$ the relation \begin{equation}\beta_{i+j}=\beta_i\cdot\beta_j^{2^i}\label{equ:betamagic}
\end{equation}
holds. By repeatedly applying Equation~(\ref{equ:betamagic}) with $i=j$, we see that computing all of $\beta_{2^0},\beta_{2^1},\dots,\beta_{2^{k_1}}$ requires no more than $\lfloor\log_2(m-1)\rfloor$ multiplications in ${\mathbb F}_{2^m}^*$ and $\lfloor\log_2(m-1)\rfloor$ exponentiations by a power of $2$. In a ghost-bit or a Gaussian normal basis representation, all occurring exponentiations are ($\alpha$-independent) permutations, and as the multiplications are of the form $\beta_j\cdot(\beta_j)^{2^j}$, to save resources we will exploit that $(\beta_j)^{2^j}$ can be derived from $\beta$---there is no need to implement a general multiplier.

Beginning with $\beta_{2^{k_1}}$, we use Equation~(\ref{equ:betamagic}) to calculate $\beta_{2^{k_1}+2^{k_2}}$ and then iterate this process to obtain $\beta_{2^{k_1}+2^{k_2}+2^{k_3}}$,  etc., until we finally reach $\beta_{m-1}=\beta_{2^{k_{1}}+2^{k_{2}}+\dots+2^{k_{\HW(m-1)}}}$. Hence, with  $\beta_{2^{k_1}},\dots,\beta_{2^{k_{\HW(m-1)}}}$ being available, $\HW(m-1)-1$ multiplications in ${\mathbb F}_{2^m}^*$ and $\HW(m-1)-1$ exponentiations by a power of $2$ suffice to derive $\beta_{m-1}$. 

\begin{example}[Itoh-Tsujii inversion]\label{exa:itotsu}
For $m=7$, we have $m-1=6=2^2+2^1$, so given an input $\alpha=\beta_{2^0}\in{\mathbb F}_{2^7}^*$, with $2\le\lfloor\log_2(6)\rfloor$ applications of Equation~\eqref{equ:betamagic} we can find $\beta_{2^1}$ and $\beta_{2^2}$. Then, with $1=\HW(6)-1$ additional application of Equation~\eqref{equ:betamagic}, we obtain $\beta_{2^2+2^1}$. After a final squaring---which in the case of a ghost-bit or a Gaussian normal basis representation is just a permutation of coefficients---yields $\alpha^{-1}=\beta_{2^2+2^1}^2$.
\end{example}

\section{Multiplying in linear depth using ghost-bit and Gaussian normal basis representations}

For implementing the inverter discussed in the sequel, the multiplication of field elements plays a crucial role. As we are interested in Gaussian normal basis and ghost-bit basis representations, we begin by detailing linear-depth circuits for multiplication in each of these representations.

\subsection{Linear depth multiplication using a ghost-bit basis}\label{sec:MastrovitoGhostbit}
To multiply two $(m+1)$-bit inputs $\ket{\alpha}$ and $\ket{\beta}$ which represent field elements $\alpha,\beta\in{\mathbb F}_{2^m}$ in a ghost-bit basis, Formula~(\ref{equ:gbbformula}) immediately yields a circuit consisting of $(m+1)^2$ Toffoli gates: each individual product $\alpha_j\beta_{(i-j)\bmod{(m+1)}}$ corresponds to a single Toffoli gate. Adopting an observation from \cite{MMCP09}, we recognize that these $(m+1)^2$ Toffoli gates can be evaluated in linear depth: for fixed $(i-2j)\bmod (m+1)$, the Toffoli gates to compute the $m+1$ products $\alpha_j\beta_{(i-j)\bmod{(m+1)}}$ $(j=0,\dots,m)$ operate on disjoint wires. Consequently, we can evaluate these $m+1$ Toffoli gates in parallel, and iterating over all $m+1$ possible values for $(i-2j)\bmod (m+1)$, we obtain a multiplier of depth $m+1$. This establishes the following result, which for the special case $\ket{\xi}=\ket{0}$ yields a basic multiplier.
\begin{proposition}
If a ghost-bit basis representation of\/ ${\mathbb F}_{2^m}$ is available, the multiplication $\ket{\alpha}\ket{\beta}\ket{\xi}\mapsto\ket{\alpha}\ket{\beta}\ket{\xi+\alpha\beta}$ with $\alpha,\beta,\xi\in{\mathbb F}_{2^m}$ can be realized in depth $m+1$ with $m^2+2m+1$ Toffoli gates.
\end{proposition}

As a concrete example of a ghost-bit basis multiplier, let us apply the above proposition to the field with $16$ elements.
\begin{example}
  Consider the ghost-bit basis representation of ${\mathbb F}_{2^4}$ from Example~\ref{exa:gbb}. In this case, evaluating all terms $\alpha_j\beta_{(i-j)\bmod 5}$ in order for $(i-2j)\bmod 5=0,1,2,3,4$ yields a multiplier of depth $5$, consisting of $5\cdot 5=25$ Toffoli gates, as shown in Figure~\ref{fig:GBBmult1}.
\begin{figure}[htb]
\caption{A ghost-bit basis multiplier for $\alpha\cdot\beta\in{\mathbb F}_{2^4}$}\label{fig:GBBmult1}
  \begin{center}
     \includegraphics[scale=0.75, trim = 60mm 145mm 140mm 25mm]{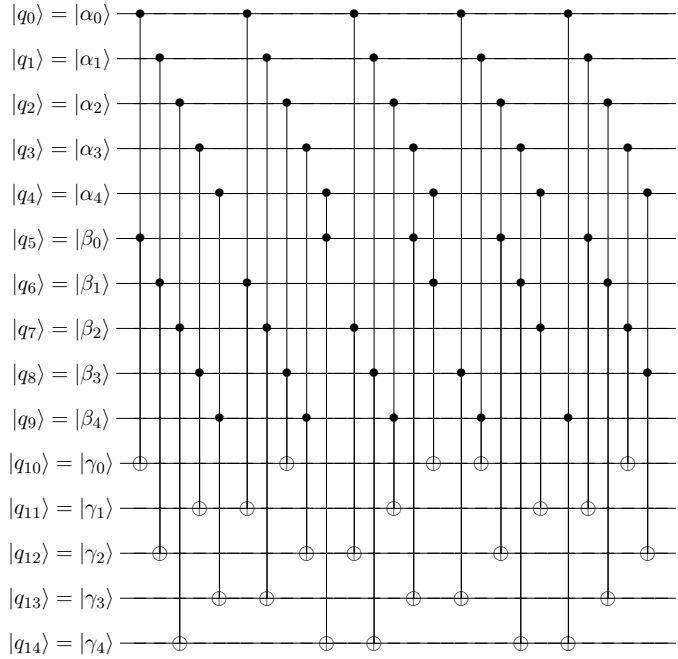}
  \end{center}
\end{figure}
\end{example}

Next, we consider the special case of computing products $\alpha\cdot\alpha^{2^j}$ with a fixed $j$, as occurring in the Itoh-Tsujii algorithm described in Section~\ref{sec:itsutsuji}. This variant of our multiplier takes as input the ghost-bit basis representation $(\alpha_0,\dots,\alpha_{m})\in{\mathbb F}_2^{m+1}$ of some $\alpha\in{\mathbb F_{2^m}}$ and a $\ket{0}$-initialized $m+1$-bit register, in which the ghost-bit basis representation $(\gamma_0, \gamma_1, \ldots, \gamma_{m})$ of $\gamma=\alpha\cdot\alpha^{2^j}$ will be stored. The total number of wires required is only $2\cdot(m+1)$.  As we are using a ghost-bit basis representation, squaring is a simple permutation, and more generally exponentiation by $2^r$ corresponds to a permutation. In particular, we can obtain the ghost-bit basis representation of $\alpha^{2^r}$ from $(\alpha_0, \alpha_1, \ldots, \alpha_{m})$ by reading out the individual entries in a different order. Hence, the following result confirms that the saving of $m$ wires can be done without sacrificing the property of having linear depth.

\begin{proposition}\label{prop:gbbparallel}
If a ghost-bit basis for ${\mathbb F}_{2^m}$ is available, then for any fixed $r\in \left\{0, \ldots, m\right\}$ the multiplication $\ket{\alpha}\ket{\xi}\mapsto\ket{\alpha}|\xi+\alpha\cdot\alpha^{2^r}\rangle$ with $\alpha,\xi\in{\mathbb F}_{2^m}$ can be realized in depth $2m+2$ using $m^2+m$ Toffoli and $m+1$ CNOT gates.
\end{proposition}

{\bf Proof:}{
Let $\alpha=\sum_{i=0}^{m}\alpha_ix^i+(x^{m+1}+1)$ be a ghost-bit basis representation for $\alpha\in{\mathbb F}_{2^m}$. Then Equation~(\ref{equ:gbbsquare}) yields $\alpha^{2^r}=\sum_{i=0}^{m}\alpha_{\pi^{-r}(i)}x^i+(x^{m+1}+1)$, and with Equation~(\ref{equ:gbbformula}) we recognize the $i^\text{th}$ coefficient of $\alpha\cdot\alpha^{2^r}$ as
$$\gamma_i=\sum_{j=0}^m\alpha_j\alpha_{\pi^{-r}((i-j)\bmod{(m+1))}}\quad(i=0,\dots,m).$$
As applying $\pi$ can be seen as doubling modulo $m+1$, applying $\pi^{-r}$ translates into division by $2^r$ modulo $m+1$. We may assume that $2^r\ne 1\bmod{(m+1)}$, as otherwise $r\in\{0,m\}$, and exponentiation with ${2^r}$ becomes the identity on ${\mathbb F}_{2^m}$. Then, for any fixed `index sum' $\sigma\in\{0,\dots,m\}$, there are exactly $m+1$ pairs $(i, j)\in\{0,\dots,m\}^2$ satisfying 
\begin{equation}
  {\pi^{-r}((i-j)\bmod{(m+1)})}+j=\sigma \bmod{(m+1)}.\label{equ:Deltaequation}
\end{equation}
Namely, for each $i\in\{0,\dots,m\}$ we obtain a unique corresponding $j\in\{0,\dots,m\}$ by solving the linear equation $$2^{-r}\cdot (i-j)+j=\sigma\bmod{(m+1)}$$ for $j$---at this we divide by ${1-2^{-r}}\pmod{m+1}$ which is possible as $2^r\ne1$. The subsequent argument shows that we can compute the $m+1$ products $\alpha_j\alpha_{\pi^{-r}((i-j)\bmod{(m+1))}}$ for those $(i,j)$-pairs satisfying Equation~\eqref{equ:Deltaequation} in depth $2$. By arranging our circuit such that the values $\sigma=0,\dots,m$ are processed in order, we achieve the claimed overall depth of $2m+2$.

Suppose we have two products $\alpha_j\alpha_{\pi^{-r}((i-j)\bmod{(m+1))}}$ and $\alpha_{j'}\alpha_{\pi^{-r}((i'-j')\bmod{(m+1))}}$ satisfying $${\pi^{-r}((i-j)\bmod{(m+1)})}+j=\sigma={\pi^{-r}((i'-j')\bmod{(m+1)})}+j',$$
then we may assume $j\ne j'$, as otherwise $${\pi^{-r}((i-j)\bmod{(m+1))}}={\pi^{-r}((i'-j')\bmod{(m+1))}},$$ and there is nothing to show. Consequently, the two gates evaluating the two terms $$\alpha_j\alpha_{\pi^{-r}((i-j)\bmod{(m+1))}}\text{ and }\alpha_{j'}\alpha_{\pi^{-r}((i'-j')\bmod{(m+1))}}$$ have different target bits. We can evaluate these two terms in parallel whenever the intersection $$\{j, \pi^{-r}((i-j)\bmod{(m+1)})\}\cap \{j', \pi^{-r}((i'-j')\bmod{(m+1)}\}$$ is empty---in this case the corresponding gates operate on disjoint wires. To better understand the situation, let us define an undirected graph $\mathfrak G$ with vertex set ${\mathbb Z}/(m+1)$, so that vertex $i+(m+1)$ corresponds to the wire representing $\alpha_{i}$. We connect two vertices, whenever they serve as control bits for the same gate, i.\,e., we include the edges
$$
\{j\bmod{(m+1)},\pi^{-r}((i-j)\bmod{(m+1)})\bmod{(m+1)}\}
$$
for all $i,j\in{\mathbb Z}/(m+1)$ with ${\pi^{-r}((i-j)\bmod{(m+1)})}+j=\sigma\bmod{(m+1)}$. In particular, we obtain exactly one self-loop ($j=\sigma/2\bmod{(m+1)}$).
Instead of using the above description of the edges, we can equivalently include all edges $$\{j\bmod{(m+1)},\sigma-j \bmod{(m+1)}\}$$
for $j\in{\mathbb Z}/(m+1)$. Because $\sigma-(\sigma-j)=j\bmod{(m+1)}$, we see that the resulting graph $\mathfrak G$ consists of $m/2$ vertex pairs, each connected by two parallel edges, and one isolated point (namely $\sigma/2 \bmod (m+1))$ with a self-loop, corresponding to a CNOT. Consequently, two colors suffice to color the edges in such a way, that no neighboring edges share a color. Now all gates corresponding to an edge with the same color operate on disjoint wires and hence can be evaluated in parallel.}\hfill $\Box$ \smallskip

To illustrate the `wire saving' offered by Proposition~\ref{prop:gbbparallel}, let us again consider the field with $16$ elements.

\begin{example}\label{ex:ghost}
   For $r=2$, the permutation $\pi^{-r}$ corresponds to a multiplication with $2^{-2}=-1\bmod{5}$, i.\,e., we have to find
$$\gamma_i\hspace*{-1.5pt}=\hspace*{-1.5pt}\alpha_0\alpha_{-i\bmod{5}}+\alpha_1\alpha_{(1-i)\bmod 5}+\alpha_2\alpha_{(2-i)\bmod 5}+\alpha_3\alpha_{(3-i)\bmod 5}+\alpha_4\alpha_{(4-i)\bmod 5}\, (i\hspace*{-1.5pt}=\hspace*{-1.5pt}0,\dots,4).$$
Using the condition $2\cdot j-i=\sigma\bmod{5}$, each of the occurring $25$ terms can be associated with a particular value of~$\sigma$:
\begin{description}
\item[$\sigma=0$:] $\alpha_0\alpha_0$, $\alpha_1\alpha_4$, $\alpha_2\alpha_3$, $\alpha_3\alpha_2$, $\alpha_4\alpha_1$
\item[$\sigma=1$:] $\alpha_0\alpha_1$, $\alpha_1\alpha_0$, $\alpha_2\alpha_4$, $\alpha_3\alpha_3$, $\alpha_4\alpha_2$
\item[$\sigma=2$:] $\alpha_0\alpha_2$, $\alpha_1\alpha_1$, $\alpha_2\alpha_0$, $\alpha_3\alpha_4$, $\alpha_4\alpha_3$
\item[$\sigma=3$:] $\alpha_0\alpha_3$, $\alpha_1\alpha_2$, $\alpha_2\alpha_1$, $\alpha_3\alpha_0$, $\alpha_4\alpha_4$
\item[$\sigma=4$:] $\alpha_0\alpha_4$, $\alpha_1\alpha_3$, $\alpha_2\alpha_2$, $\alpha_3\alpha_1$, $\alpha_4\alpha_0$
\end{description}
The resulting graph for $\sigma=0$ is shown in Figure \ref{fig:ghost}. 

\begin{figure}[hbt]
\caption{\label{fig:ghost}  Graph representing the term for $\sigma=0$ as described in Example \ref{ex:ghost}. The $2$-coloring of 
the edges---where the different line styles indicate the colors---translates into a depth $2$ circuit for this term. 
}
\begin{center}
\begin{tikzpicture}
  \tikzstyle{every node}=[draw,shape=circle];
  \tikzset{every loop/.style={}}
  \path (90+0*72:1.5cm) node (P0) {$0$};
  \path (90+1*72:1.5cm) node (P1) {$1$};
  \path (90+2*72:1.5cm) node (P2) {$2$};
  \path (90+3*72:1.5cm) node (P3) {$3$};
  \path (90+4*72:1.5cm) node (P4) {$4$};

  \path (P0) edge [loop right] (P0);
  \path[dashed] (P1) edge [bend right] (P4);
  \path (P4) edge [bend right] (P1);
  \path[dashed] (P2) edge [bend right] (P3);
  \path (P3) edge [bend right] (P2);

\end{tikzpicture}
\end{center}
\end{figure}
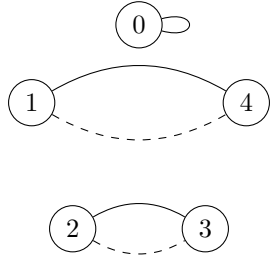

 Each edge corresponds to one gate, and with the 2-coloring of the edges we obtain a depth~$2$ circuit for evaluating the terms associated with $\sigma=0$ (and add them to the respective input/partial result $\gamma_i$). Applying a similar reasoning to the other $\sigma$-values, we obtain a circuit of depth $10$ for implementing the map $\ket{\alpha}\ket{\xi}\mapsto\ket{\alpha}\ket{\xi+\alpha\cdot\alpha^4}$ for $\alpha,\xi\in{\mathbb F}_{2^4}$, as seen in Figure~\ref{fig:GBBmult2}.
\begin{figure}[htb]
\caption{A ghost-bit basis multiplier for $\alpha\cdot\alpha^{2^2}\in{\mathbb F}_{2^4}$}\label{fig:GBBmult2}
  \begin{center}
     \includegraphics[scale=0.75, trim = 40mm 185mm 80mm 25mm]{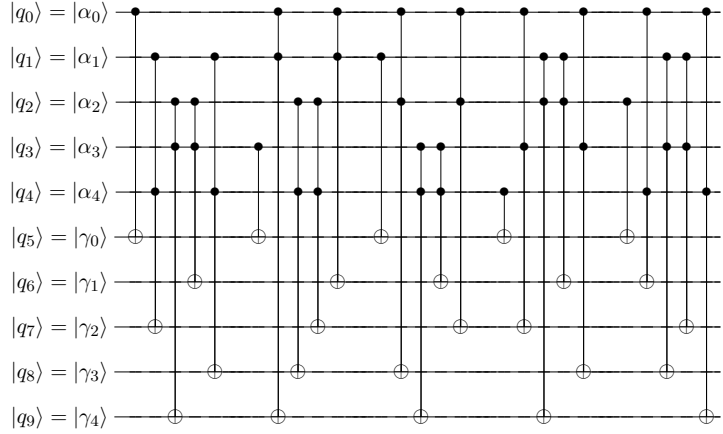}
  \end{center}
\end{figure}
\end{example}

\subsection{Linear depth multiplication using a Gaussian normal basis}
Assume ${\mathbb F_{2^m}}$ has a Gaussian normal basis of type $t$. Our multiplier takes as input the normal basis representations $(\alpha_0, \alpha_1, \ldots, \alpha_{m-1})\in{\mathbb F}_2^m$ and $(\beta_0, \beta_1, \ldots, \beta_{m-1})\in{\mathbb F}_2^m$ of two elements $\alpha,\beta\in{\mathbb F}_{2^m}$, along with a $\ket{0}$-initialized $m$-bit register, in which the normal basis representation $(\gamma_0, \gamma_1, \ldots, \gamma_{m-1})$ of $\gamma=\alpha\cdot\beta$ will be stored. Consequently, the total number of wires is $3m$. Each coefficient product $\alpha_j\beta_k$ in Equation~(\ref{equ:gnbmult}) can be realized with a Toffoli gate, and so for a fixed $i\in\{0,\dots,m-1\}$ we can compute $\gamma_i$ with at most
$$
\left\{\begin{array}{ll}tm-1\text{ consecutive Toffoli gates}&\text{, if $t$ is even}\\
                        tm-1+2\cdot(m/2)=(t+1)m-1\text{ consecutive Toffoli gates}&\text{, if $t$ is odd}\\
\end{array}
\right..
$$
From this we immediately obtain an overall gate count of $(t+(t\bmod 2))\cdot m^2-m$ Toffoli gates for our normal basis multiplier. This multiplier can be realized in linear depth: fix an arbitrary $k\in\{1,\dots,tm-1\}$ and two different positions $i,i'\in\{0,\dots,m-1\}$ in the normal basis representation of the product $\gamma=\alpha\cdot\beta$. Then the Toffoli gates computing $\alpha_{F(k+1)+i}\beta_{F(p-k)+i}$ and $\alpha_{F(k+1)+i'}\beta_{F(p-k)+i'}$ operate on disjoint wire sets, as obviously
\begin{eqnarray*}{F(k+1)+i}&\ne&{F(k+1)+i'}\pmod m\text{ and}\\
{F(p-k)+i}&\ne&{F(p-k)+i'}\pmod m.
\end{eqnarray*}
For odd $t$, we see analogously that $\alpha_{k-1+i}\beta_{k-1+\frac{m}{2}+i}$ can be calculated in parallel with $\alpha_{k-1+i'}\beta_{k-1+\frac{m}{2}+i'}$ for all $i\ne i'$, and $\alpha_{k-1+\frac{m}{2}+i}\beta_{k-1+i}$ can be calculated in parallel with $\alpha_{k-1+\frac{m}{2}+i'}\beta_{k-1+i'}$ for all $i\ne i'$, as summarized in the following result.
\begin{proposition}
If a Gaussian normal basis of type $t$ is available for ${\mathbb F}_{2^m}$, the multiplication $\ket{\alpha}\ket{\beta}\ket{\xi}\mapsto\ket{\alpha}\ket{\beta}\ket{\xi+\alpha\beta}$ of two field elements $\alpha,\beta\in{\mathbb F}_{2^m}$ can be realized in depth $(t+(t\bmod 2))\cdot m-1$ using $(t+(t\bmod 2))\cdot m^2-m$ Toffoli gates.
\end{proposition}
As a concrete example of a Gaussian normal basis multiplier, let us apply the above proposition to the field with $32$ elements.
\begin{example}\label{exa:gnbmultgeneral}
  Consider the type~2 Gaussian normal basis from Example~\ref{exa:gnb}. Here the product $\gamma=\alpha\cdot\beta$ of $\alpha,\beta\in{\mathbb F}_{2^5}$, is represented by $(\gamma_0,\dots,\gamma_{m-1})$ with
\begin{eqnarray*}\gamma_i&=&\alpha_{1+i}\beta_i + \alpha_{3+i}\beta_{1+i} +\alpha_{2+i}\beta_{3+i} +\alpha_{4+i}\beta_{2+i} +\alpha_{4+i}\beta_{4+i} +\nonumber\\&&\alpha_{2+i}\beta_{4+i} +\alpha_{3+i}\beta_{2+i} +\alpha_{1+i}\beta_{3+i} +\alpha_i\beta_{1+i}.\end{eqnarray*}
Implementing this summation term by term yields a normal basis multiplier for ${\mathbb F}_{2^5}$ comprised of $9\cdot 5=45$ Toffoli gates and of total depth~9 (each term of the summation can be evaluated in parallel for $i=0,\dots,4$), as seen in Figure~\ref{fig:GNBmult1}.

\begin{figure}[htb]
\caption{A Gaussian normal basis multiplier for $\alpha\cdot\beta\in {\mathbb F}_{2^5}$}\label{fig:GNBmult1}
  \begin{center}
     \includegraphics[scale=0.75, trim = 40mm 145mm 40mm 25mm]{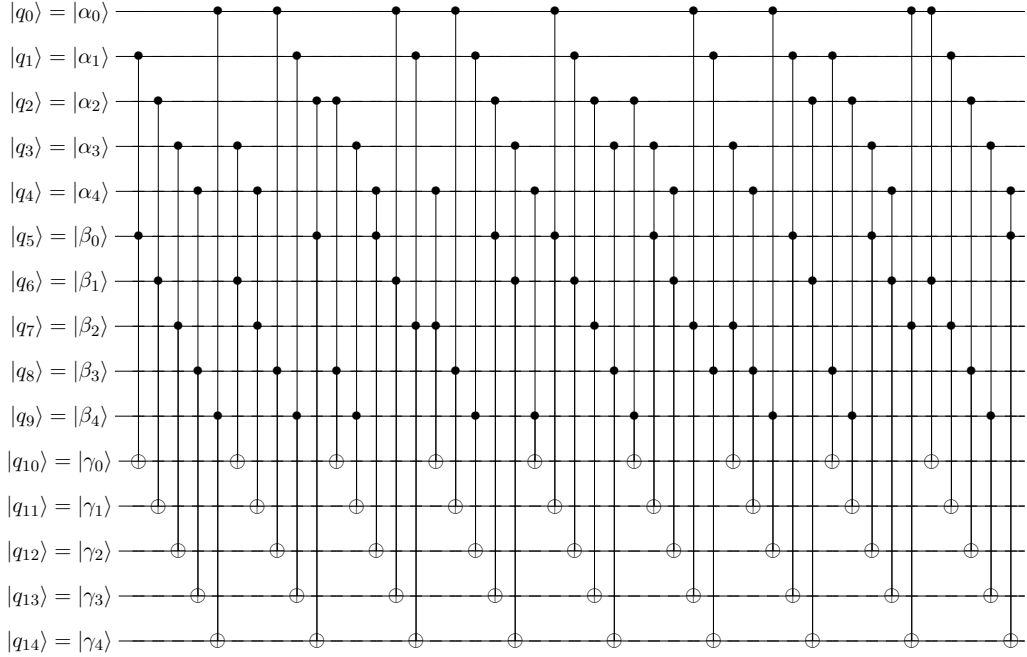}
  \end{center}
\end{figure}

\end{example}

Similarly, as in the case of a ghost-bit basis representation, it is possible to compute products of the form $\alpha\cdot\alpha^{2^r}$ in linear depth without having $\alpha^{2^r}$ represented as a separate input. Hence, the following result shows that the saving of $m$ wires can be done without sacrificing the property of having linear depth.

\begin{proposition}\label{prop:gnbparallel}
If a Gaussian normal basis of type $t$ is available for ${\mathbb F}_{2^m}$, for any fixed $r\in \left\{0, \ldots, m\right\}$ the multiplication $\ket{\alpha}\ket{\xi}\mapsto\ket{\alpha}|\xi+\alpha\cdot\alpha^{2^r}\rangle$ for $\alpha\in{\mathbb F}_{2^m}$ can be realized in depth $3\cdot (t+(t\bmod 2))\cdot m-3$ using $(t+(t\bmod 2))\cdot m^2-m$ gates (CNOT or Toffoli).
\end{proposition}

{\bf Proof:}
Using Equation~(\ref{equ:gnbmult}) to calculate the product $\alpha\cdot\alpha^{2^j}$ again, the upper bound for the total number of gates remains unchanged. It could happen, however, that the control bits of a Toffoli gate end up on the same wire, so that instead of a Toffoli we obtain a CNOT gate.

To argue that the circuit depth grows at most by a factor of $3$, we fix $k\in\{1,\dots,tm-1\}$ arbitrary. Then $\beta_k = \alpha_{k-r}$, and we claim that all $m$ terms \begin{equation}\alpha_{F(k+1)+i}\beta_{F(p-k)+i}=\alpha_{F(k+1)+i}\alpha_{F(p-k)-r+i}\quad (i=0, \ldots, m-1)\label{equ:onegnblayer}
\end{equation} can be calculated in parallel using depth at most 3.
\begin{description}
\item[Case $F(k+1)=F(p-k)-r\ (\bmod m)$:] Here, instead of Toffoli gates, we have only CNOT gates operating on disjoint wires. Hence, all $m$ terms can be computed at the same time, i.,\,e., in depth~1.
\item[Case $F(k+1)\ne F(p-k)-r\ (\bmod m)$:] For $i\ne i'$, we can evaluate the terms $$\alpha_{F(k+1)+i}\alpha_{F(p-k)-r+i}\text{ and }\alpha_{F(k+1)+i'}\alpha_{F(p-k)-r+i'}$$ in parallel whenever the two sets $$\left\{{F(k+1)+i, F(p-k)-r+i}\right\}\text{ and }\left\{{F(k+1)+i', F(p-k)-r+i'}\right\}$$ have an empty intersection, meaning the two Toffoli gates operate on disjoint wires. We define an undirected graph $\mathfrak G$ with vertex set ${\mathbb Z}/(m)$---so vertex $i+(m)$ corresponds to the wire representing $\alpha_{i\ (\bmod m)}$---and edge set $$E:=\left\{{\{F(k+1)+i\ (\bmod m), F(p-k)-r+i\ (\bmod m)\}}:i=0, \ldots, m-1\right\},$$
\end{description}
i.\,e., each edge corresponds to one Toffoli gate. If we can find an edge coloring of this graph such that neighboring edges always have different colors, then all Toffoli gates corresponding to the same color can be calculated in parallel. We show that 3 colors will be sufficient, and hence a depth 3 circuit suffices to compute all the products in~(\ref{equ:onegnblayer}). For $\delta=F(p-k)-r-F(k+1)$, let $\langle\delta\rangle$ be the cyclic subgroup generated by $\delta+(m)$ in ${\mathbb Z}/(m)$, and let \begin{equation}{\mathbb Z}/(m)=G_1\uplus\dots\uplus G_t\label{equ:graphdecomp}
\end{equation}
be the decomposition of ${\mathbb Z}/(m)$ into $\langle\delta\rangle$-cosets. Rewriting the edge set $E$ as $$E=\left\{\{i\ (\bmod m), i+\delta\ (\bmod m)\}|i\in\{0, \ldots, m-1\}\right\},$$ we see that the decompositon~(\ref{equ:graphdecomp}) actually yields a decomposition of the graph $\mathfrak G$---there are no edges between vertices in $G_j$ and $G_{j'}$ if $j\ne j'$. Moreover, $\langle \delta\rangle $ is cyclic with generator $\delta+(m)$, so for each $G_j$, the subgraph of $\mathfrak G$ with vertex set $G_j$ is a closed cycle on $\ord{(\delta+(m))}$ vertices. As such, we may alternatively color the edges in such a cycle red and blue. Then neighboring edges can only obtain the same color at the very last step when we try to close the cycle---this happens whenever $\ord{(\delta+(m))}$ is odd. Hence, for the last edge in a cycle, a third color may be needed. As there are no edges between the individual cycles, we have found the desired 3-coloring of $E$.

The above argument takes care of all even $t$-values, and for odd $t$-values the first of the summations in Equation~(\ref{equ:gnbmult}) is taken care of as well.\footnote{For $\ord{(\delta+(m))}=2$ the sets $G_j$ consist of two vertices, and we actually face graphs with a  $2$-coloring of the edges.} To argue that for fixed $k$ the terms $\alpha_{k-1+i}\alpha_{k-1+\frac{m}{2}-r+i}$ $(i=1,\dots,m)$ and $\alpha_{k-1+\frac{m}{2}+i}\alpha_{k-1-r+i}$ ($i=1,\dots,m)$ can be computed in depth 3, we can use an analogous argument as above, replacing $\delta$ with $(m/2)-r$ and $(m/2)+r$, respectively.\hfill$\square$\smallskip


\begin{example}\label{ex:gaussian}
  Sticking with the Gaussian normal basis representation of ${\mathbb F}_{2^5}$ from Example~\ref{exa:gnbmultgeneral}, let us consider the special case of a multiplication $\gamma=\alpha\cdot\beta$ where $\beta=\alpha^{2^1}$, i.\,e., $r=1$. Then we have $\beta_k=\alpha_{k-1}$ and Equation~(\ref{exa:gnbbeta}) can be rewritten as $\gamma_i=$
$$\underline{\alpha_{1+i}\alpha_{4+i}} + \alpha_{3+i}\alpha_{i} +\alpha_{2+i}\alpha_{2+i} +\underline{\alpha_{4+i}\alpha_{1+i}} +\alpha_{4+i}\alpha_{3+i} +\alpha_{2+i}\alpha_{3+i} +\alpha_{3+i}\alpha_{1+i} +\alpha_{1+i}\alpha_{2+i} +\alpha_i\alpha_{i}.$$
In particular, the addition of the terms $\alpha_{2+i}\alpha_{2+i}$ and $\alpha_i\alpha_{i}$ can be implemented with CNOT instead of Toffoli gates, fulfilling the condition $F(k+1)=F(11-k)-1$ as in the first case of the above proof. We also note that the underlined terms cancel each other, which yields a simplification of our circuit that is not reflected by the upper bounds in Proposition~\ref{prop:gnbparallel}.

Going through the remaining values for $k$ (for which $F(k+1)\ne F(11-k)-1$ and no cancellation occurs), we obtain the following values $\delta=F(11-k)-1-F(k+1)$:

$$\begin{array}{c|c|c|c|c|c}
k&2&5&6&7&8\\\hline
\delta&-3&-1&1&-2&1
\end{array}
$$
As $m=5$ is prime, each $\delta+(5)$ generates the complete additive group ${\mathbb Z}/(5)$, and so the graph $\mathfrak G$ is simply a closed cycle. For instance, consider $k=5$ such that $\delta=-1$. Then the graph in Figure~\ref{fig:penta} is obtained, where a vertex labeled $i$ $(i=0,\dots,4)$ represents the residue class $i+(5)$, and different line styles indicate different colors.
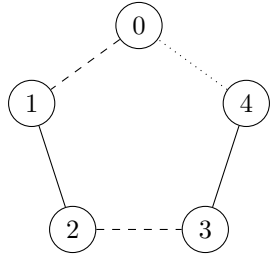
\begin{figure}[hbt]
\caption{\label{fig:penta} Graph corresponding to the cosets $\delta+(5)$ for $\delta=-1$ 
as described in Example \ref{ex:gaussian}. The $3$-coloring of 
the edges---where the different line styles in the pentagon indicate the three 
different colors---translates into a depth $3$ circuit. 
}
\begin{center}
\begin{tikzpicture}
  \tikzstyle{every node}=[draw,shape=circle];
  \path (90+0*72:1.5cm) node (P0) {$0$};
  \path (90+1*72:1.5cm) node (P1) {$1$};
  \path (90+2*72:1.5cm) node (P2) {$2$};
  \path (90+3*72:1.5cm) node (P3) {$3$};
  \path (90+4*72:1.5cm) node (P4) {$4$};

  \draw[dashed] (P0) -- (P1);
  \draw         (P1) -- (P2);
  \draw[dashed] (P2) -- (P3);
  \draw         (P3) -- (P4);
  \draw[dotted] (P4) -- (P0);
\end{tikzpicture}
\end{center}
\end{figure}

As shown in Figure~\ref{fig:GNBmult2}, this 3-coloring translates into a quantum circuit  of depth~3 to compute the terms $\alpha_{4+i}\alpha_{3+i}$ $(i=0,\dots,4)$ (and add them to the respective input/partial result $\gamma_i$).
\begin{figure}[htb]
\caption{Part of a Gaussian normal basis multiplier for $\alpha\cdot\alpha^{2^1}\in {\mathbb F}_{2^5}$: computing the terms $\alpha_{4+i}\alpha_{3+i}$}\label{fig:GNBmult2}
  \begin{center}
     \includegraphics[scale=0.75, trim = 40mm 180mm 180mm 25mm]{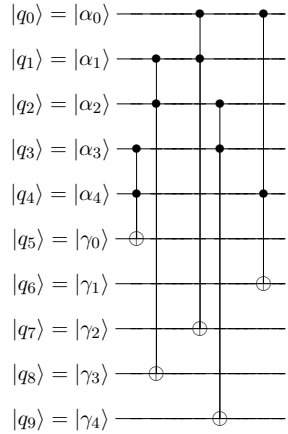}
  \end{center}
\end{figure}
\end{example}

\section{Inversion in depth {\boldmath$\bigO(m\log(m))$ using the Itoh-Tsujii algorithm}}
With the linear depth multipliers from the previous section, we can now implement a depth $\bigO(m\log(m))$ algorithm to invert field elements $\alpha\in{\mathbb F}_{2^m}^*$, if a Gaussian normal basis or ghost-bit basis representation is available.

The first part of the input is, respectively, an $m$ or $(m+1)$-bit representation $\ket{\alpha}$ of the element $\alpha\in{\mathbb F}_{2^m}^*$ to be inverted.\footnote{The input $\ket{0}$ for $\ket{\alpha}$ results in the output $\ket{0}$ as `inverse.'} Now, providing $\lfloor\log_2(m-1)\rfloor$ auxiliary registers that are initialized with $\ket{0}$, a sequence of ${\lfloor\log_2(m-1)\rfloor}$ consecutive multipliers can be used to calculate the values $\beta_{2^0},\beta_{2^1},\dots,\beta_{2^{k_1}}$ from Section~\ref{sec:itsutsuji}---recall that $\beta_{2^0}=\alpha$. From Proposition~\ref{prop:gbbparallel} and Proposition~\ref{prop:gnbparallel}, we obtain the following resource counts for this part of the inverter computation:

\begin{itemize}
  \item If a ghost-bit basis representation of ${\mathbb F}_{2^m}$ is available, we can find all of $\beta_{2^0},\beta_{2^1},\dots,\beta_{2^{k_1}}$ in depth $\lfloor\log_2(m-1)\rfloor\cdot (2m+2)$ using $\lfloor\log_2(m-1)\rfloor\cdot(m^2+m)$ Toffoli and $\lfloor\log_2(m-1)\rfloor\cdot(m+1)$ CNOT gates. In doing so, $(1+\lfloor\log_2(m-1)\rfloor)\cdot(m+1)$~qubits suffice.
  \item Assume that a Gaussian normal basis representation of ${\mathbb F}_{2^m}$ is available. Then we can find all of $\beta_{2^0},\beta_{2^1},\dots,\beta_{2^{k_1}}$ in depth $\lfloor\log_2(m-1)\rfloor\cdot(3\cdot(t+(t\bmod 2))\cdot m-3)$ using $\lfloor\log_2(m-1)\rfloor\cdot((t+(t\bmod 2))\cdot m^2-m)$ gates (CNOT or Toffoli). In doing so, $(1+\lfloor\log_2(m-1)\rfloor)\cdot m$~qubits suffice.
\end{itemize}
At this point, our inverter has computed all of $\beta_{2^0},\beta_{2^1},\dots,\beta_{2^{k_1}}$ and stored each of these values in a separate set of wires. Next, we can use a sequence of $\HW(m-1)-1$ (general) multipliers, each obtaining an auxiliary input $\ket{0}$, to gather the actually needed values $\beta_{2^{k_1}},\dots,\beta_{2^{k_{\HW(m-1)}}}$ and form their product using Equation~\eqref{equ:betamagic}. All exponentiations of the form $\beta_j^{2^i}$ are for free, in that a multiplier can just read out the coefficients of the respective $\beta_j$ in permuted order to obtain the required input value. This is simply a permutation of the control bit positions. Consequently, we have the following resource counts:
\begin{itemize}
\item If a ghost-bit basis of ${\mathbb F}_{2^m}$ is available and given $|{\beta_{2^{k_1}}}\rangle,\dots,|{\beta_{2^{k_{\HW(m-1)}}}}\rangle$, we can compute $\ket{\beta_{m-1}}$ in depth $(\HW(m-1)-1)\cdot(m+1)$ using $(\HW(m-1)-1)\cdot(m^2+2m+1)$ Toffoli gates. For the auxiliary inputs $\ket{0}$ respectively storing some intermediate results, $(\HW(m-1)-1)\cdot(m+1)$~qubits suffice.
\item If a Gaussian normal basis of ${\mathbb F}_{2^m}$ is available and given $|{\beta_{2^{k_1}}}\rangle,\dots,|{\beta_{2^{k_{\HW(m-1)}}}}\rangle$, we can compute $\ket{\beta_{m-1}}$ in depth $(\HW(m-1)-1)\cdot((t+(t\bmod 2))\cdot m-1)$ using $(\HW(m-1)-1)\cdot((t+(t\bmod 2))\cdot m^2-m)$ Toffoli gates. For the auxiliary inputs $\ket{0}$ respectively storing some intermediate results, $(\HW(m-1)-1)\cdot m$~qubits are needed.
\end{itemize}
The final squaring operation in the Itoh-Tsujii algorithm is again for free, in that the last multiplier can simply write out the result in permuted order. In summary, we obtain the following estimate for a ghost-bit basis, where we double depth and gate count to account for the resources to `uncompute' auxiliary values---this is an upper bound, as the last multiplication actually does not have to be `undone.'

\begin{proposition}\label{prop:ghost}
If a ghost-bit basis for ${\mathbb F}_{2^m}$ is available, the inversion $\ket{\alpha}\ket{0}\mapsto|{\alpha^{-1}}\rangle\ket{0}$ can be implemented in depth $2\cdot\lfloor\log_2(m-1)\rfloor\cdot(2m+2) + 2\cdot(\HW(m-1)-1)\cdot(m+1)=\bigO(m\log_2(m))$ and using $2\cdot\lfloor\log_2(m-1)\rfloor\cdot(m^2+m) + 2\cdot(\HW(m-1)-1)\cdot(m^2+2m+1)$ Toffoli and $2\cdot \lfloor\log_2(m-1)\rfloor\cdot(m+1)$ CNOT gates. The inversion can be implemented with $(1+\lfloor\log_2(m-1)\rfloor)\cdot(m+1)+(\HW(m-1)-1)\cdot(m+1)=\bigO(m\log_2(m))$~qubits.
\end{proposition}
Analogously, adding the respective bounds for the case of a Gaussian normal basis of type $t$ yields the following estimate. If we consider $t$ as constant, the depth of the resulting circuit is again in $\bigO(m\log_2(m))$.

\begin{proposition}\label{prop:gaussian}
If a Gaussian normal basis of type $t$ for ${\mathbb F}_{2^m}$ is available, the inversion $\ket{\alpha}\ket{0}\mapsto|{\alpha^{-1}}\rangle\ket{0}$ can be implemented in depth $\lfloor\log_2(m-1)\rfloor\cdot(6\cdot(t+(t\bmod 2))\cdot m-6) + 2\cdot(\HW(m-1)-1)\cdot((t+(t\bmod 2))\cdot m-1)=\bigO(m\log_2(m))$ using $2\cdot \lfloor\log_2(m-1)\rfloor\cdot((t+(t\bmod 2))\cdot m^2-m) + 2\cdot(\HW(m-1)-1)\cdot((t+(t\bmod 2))\cdot m^2-m)$ gates (CNOT or Toffoli). The inversion can be implemented with $(1+\lfloor\log_2(m-1)\rfloor)\cdot m+(\HW(m-1)-1)\cdot m=\bigO(m\log_2(m))$~qubits.
\end{proposition}

It is worth noting that if our extension degree $m$ has the form $m=2^n+1$, e.\,g., for $m$ being a Fermat prime, the Hamming weight of $m-1$ is one, i.\,e., we can restrict to special multipliers as described in Proposition~\ref{prop:gbbparallel} and Proposition~\ref{prop:gnbparallel} entirely. As in the general case, the last multiplier can output the result $\beta_{2^{m-1}}$ in permuted order, so that the correct inverse $(\beta_{m-1})^2$ is obtained without the need to implement a squaring operation.

Avoiding such a special case, the following example illustrates the structure of the discussed inverter with an extension of degree $7$, where a general multiplier with two arguments is brought to use.

\begin{example}
  Consider the field ${\mathbb F}_{2^7}$ we discussed in Example~\ref{exa:itotsu}, and assume a Gaussian normal basis representation is used. Then, to compute $\alpha^{-1}$ from an input $\alpha=\beta_{2^0}\in{\mathbb F}_{2^7}^*$, we can use two special multipliers as described in Proposition~\ref{prop:gnbparallel} to compute $\beta_{2^1}$ and $\beta_{2^2}$. Interpreting the input wires in appropriately permuted order, one general multiplier suffices to compute $\beta_{2^2+2^1}$. In addition, writing the output in appropriately permuted order, the output of this multiplier is actually $\beta_{2^2+2^1}^2$.

Representing an $m$-qubit input by a single wire, the structure of the resulting inverter in ${\mathbb F}_{2^7}^*$ is summarized below in Figure~\ref{fig:invert}. 

\begin{figure}[htb]
\caption{A ghost-bit or Gaussian normal basis inverter for $\alpha\in{\mathbb F}_{2^7}^*$}\label{fig:invert}
  \begin{center}
     \includegraphics[scale=0.75, trim = 70mm 230mm 170mm 25mm]{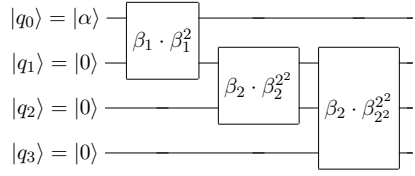}
  \end{center}
\end{figure}
\end{example}

Finally, we obtain as direct consequence of Propositions~\ref{prop:ghost} and \ref{prop:gaussian} the following 
corollary which gives an upper bound on the number of $T$-gates to perform inversion in a binary finite field where a ghost-bit basis or a Gaussian normal basis representation is available. This is a straightforward 
consequence of a realization \cite[Chapter 4.2]{NC:2000} of a Toffoli gate using $7$ $T$-gates 
(or $T^\dagger$-gates which we assume to have the same cost) in a circuit of overall $T$-depth of $6$.

\begin{corollary}
If a ghost-bit basis for ${\mathbb F}_{2^m}$ is available, an inverter can 
be implemented with a $T$-depth of at most $12\cdot\lfloor\log_2(m-1)\rfloor\cdot(2m+2) + 12\cdot(\HW(m-1)-1)\cdot(m+1)$ and using no more than $14\cdot\lfloor\log_2(m-1)\rfloor\cdot(m^2+m) + 14\cdot(\HW(m-1)-1)\cdot(m^2+2m+1)$ many $T$-gates.

If a Gaussian normal basis of type $t$ for ${\mathbb F}_{2^m}$ is available, an inverter can be implemented with a $T$-depth of at most  $6\cdot \lfloor\log_2(m-1)\rfloor\cdot(6\cdot(t+(t\bmod 2))\cdot m-6) + (12\cdot \HW(m-1)-6)\cdot((t+(t\bmod 2))\cdot m-1)$ using at most $14\cdot \lfloor\log_2(m-1)\rfloor\cdot((t+(t\bmod 2))\cdot m^2-m) + 14\cdot(\HW(m-1)-1)\cdot((t+(t\bmod 2))\cdot m^2-m)$ many $T$-gates. 
\end{corollary}

\section{Comparison and conclusions}
The above discussion demonstrates that the use of representations of finite fields other than a polynomial basis, can enable efficient and elegant quantum circuits for realizing binary finite field arithmetic. Table~\ref{tab:depthcomparison} gives a brief asymptotic comparison of the circuit depth of the representations discussed here in comparison to a polynomial basis representation. For a Gaussian normal basis representation the exact depth increases when the type $t$ gets larger, but for cryptographic purposes already a value of $t=10$ is unusually high, and small values like $t=2$ or $t=4$ are more typical; here, we consider $t$ as a (small) constant.

\begin{center}
\begin{table}[htb]
\caption{Circuit depth of ${\mathbb F}_{2^m}$-operations for different representations}\label{tab:depthcomparison}
\begin{center}
\begin{tabular}{c|c|c|c}
& Addition & Multiplication & Inversion\\
\hline
polynomial basis \cite{BBF03,MMCP09,KaZa04}&$\bigO(1)$&$\bigO(m)$&ext. Euclidean alg.: $\bigO(m^2)$\\\hline
ghost-bit basis&$\bigO(1)$&$\bigO(m)$&Itoh-Tsujii alg.: $\bigO(m\log(m))$\\\hline
Gaussian normal basis&$\bigO(1)$&$\bigO(m)$&Itoh-Tsujii alg.: $\bigO(m\log(m))$
\end{tabular}
\end{center}
\end{table}
\end{center}

Table~\ref{tab:gatecomparison} gives an asymptotic comparison for the number of gates involved. Again, for Gaussian normal bases we consider the type $t$ as a (small) constant.

\begin{center}
\begin{table}[htb]
\caption{Number of gates when implementing ${\mathbb F}_{2^m}$-operations for different representations}\label{tab:gatecomparison}
\begin{center}
\begin{tabular}{c|c|c|c}
& Addition & Multiplication & Inversion\\
\hline
polynomial basis \cite{BBF03,MMCP09,KaZa04}&$\bigO(m)$&$\bigO(m^2)$&$\bigO(m^3)$\\\hline
ghost-bit basis&$\bigO(m)$&$\bigO(m^2)$&$\bigO(m^2\log(m))$\\\hline
Gaussian normal basis&$\bigO(m)$&$\bigO(m^2)$&$\bigO(m^2\log(m))$
\end{tabular}
\end{center}
\end{table}
\end{center}

Overall, a main feature of the presentations considered here is the convenient implementation of inversion in ${\mathbb F}_{2^m}$: having available a `free' squaring operation, the discussed technique by Itoh and Tsujii offers a viable alternative to Euclid's algorithm. It appears worthwhile to further explore the potential of different finite field representations for deriving quantum circuits that can, e.\,g., be used in connection with Shor's algorithm. From a cryptographic point of view, binary fields certainly play a prominent role, but, e.g., the discussion of \emph{Optimal Extension Fields} by Bailey and Paar \cite{BaPa98} illustrates that finite fields of larger characteristic are of cryptographic interest as well, as they can facilitate efficient (classical) implementations. Exploring different representations of finite fields with odd characteristic appears to be a worthwhile endeavor for future work.

\section*{Acknowledgments}
\noindent
BA and RS acknowledge support by NSF grant No.~1049296 \emph{(Small-scale Quantum Circuits with Applications in Cryptanalysis)}. MR acknowledges support by the Intelligence Advanced Research Projects Activity 
(IARPA) via Department of Interior National Business Center contract No.~D11PC20166. The U.S. Government is authorized to reproduce and distribute reprints for Governmental purposes notwithstanding any copyright annotation thereon. Disclaimer: The views and conclusions contained herein are those of the authors and should not be interpreted as necessarily representing the official policies or endorsements, either expressed or implied, of IARPA, DoI/NBC, or the U.S. Government.
 The authors also would like to thank the anonymous reviewers for helpful comments.

\noindent
\nocite{WCP97}
\bibliographystyle{plain}
\bibliography{GF2short}

\begin{thebibliography}{10}

\bibitem{BaPa98}
D.~V. Bailey and C.~Paar.
\newblock {Optimal Extension Fields for Fast Arithmetic in Public-Key
  Algorithms}.
\newblock In Hugo Krawczyk, editor, {\em Advances in Cryptology -- CRYPTO '98},
  volume 1462 of {\em Lecture Notes in Computer Science}, pages 472--485.
  Springer, 1998.

\bibitem{DCW:2002}
J.~N.~de Beaudrap, R.~Cleve, and J.~Watrous.
\newblock {Sharp Quantum versus Classical Query Complexity Separations}.
\newblock {\em Algorithmica}, 34(4):449--461, 2002.

\bibitem{BBF03}
S.~Beauregard, G.~Brassard, and J.~M. Fernandez.
\newblock {Quantum Arithmetic on Galois Fields}.
\newblock arXiv:quant-ph/0301163v1, January 2003.
\newblock Available at \url{http://arxiv.org/abs/quant-ph/0301163v1}.

\bibitem{WCP97}
W.~Bosma, J.~Cannon, and C.~Playoust.
\newblock {The Magma algebra system. I. The user language}.
\newblock {\em Journal of Symbolic Computation}, 24:235--265, 1997.

\bibitem{CSV:2007}
A.~M. Childs, L.~J. Schulman, and U.~V. Vazirani.
\newblock Quantum algorithms for hidden nonlinear structures.
\newblock In {\em Proceedings of the 48th Annual IEEE Symposium on Foundations
  of Computer Science (FOCS'07)}, pages 395--404. IEEE Computer Society, 2007.

\bibitem{DHH06}
R.~Dahab, D.~Hankerson, F.~Hu, M.~Long, J.~L{\'o}pez, and A.~Menezes.
\newblock {Software Multiplication Using Gaussian Normal Bases}.
\newblock {\em IEEE Transactions on Computers}, 55(8), 2006.

\bibitem{vDHI:2003}
{W. van} Dam, S.~Hallgren, and L.~Ip.
\newblock Quantum algorithms for some hidden shift problems.
\newblock In {\em {Proceedings of the 14th Annual ACM-SIAM Symposium on
  Discrete Algorithms (SODA'03)}}, pages 489--498, 2003.
\newblock Available at \url{http://arxiv.org/abs/quant-ph/0211140v1}.

\bibitem{ASG:2009}
A.~G. Fowler, A.~M. Stephens, and P.~Groszkowski.
\newblock High threshold universal quantum computation on the surface code.
\newblock {\em Phys.~Rev.~A}, 80:052312, 2009.

\bibitem{GeLu03}
W.~Geiselmann and H.~Lukhaub.
\newblock {Redundant Representation of Finite Fields}.
\newblock In Kwangjo Kim, editor, {\em Public Key Cryptography, 4th
  International Workshop on Practice and Theory in Public Key Cryptography, PKC
  2001}, volume 1992 of {\em Lecture Notes in Computer Science}, pages
  339--352. Springer, 2001.

\bibitem{Gua11}
J.~Guajardo.
\newblock {Itoh-Tsujii Inversion Algorithm}.
\newblock In Henk C.~A. van Tilborg and Sushil Jajodia, editors, {\em
  Encyclopedia of Cryptography and Security}, pages 650--653. Springer, second
  edition, 2011.

\bibitem{Hallgren:2002}
S.~Hallgren.
\newblock {Polynomial-time quantum algorithms for Pell's equation and the
  principal ideal problem}.
\newblock In {\em Proceedings of the 34th Annual ACM Symposium on Theory of
  Computing (STOC'02)}, pages 653--658, 2002.

\bibitem{ItTs89}
T.~Itoh and S.~Tsujii.
\newblock {Structure of parallel multipliers for a class of fields $GF(2^m)$}.
\newblock {\em Information and Computation}, 83:21--40, 1989.

\bibitem{JMV01}
D.~Johnson, A.~Menezes, and S.~Vanstone.
\newblock {The Elliptic Curve Digital Signature Algorithm (ECDSA)}.
\newblock {\em International Journal of Information Security}, 1(1):36--63,
  2001.

\bibitem{Jun93}
D.~Jungnickel.
\newblock {\em Finite Fields: Structure and Arithmetics}.
\newblock Wissenschaftsverlag, 1993.

\bibitem{KaZa04}
P.~Kaye and C.~Zalka.
\newblock {Optimized quantum implementation of elliptic curve arithmetic over
  binary fields}.
\newblock arXiv:quant-ph/0407095v1, July 2004.
\newblock Available at \url{http://arxiv.org/abs/quant-ph/0407095v1}.

\bibitem{Kitaev:97}
A.~Y. Kitaev.
\newblock Quantum computations: algorithms and error correction.
\newblock {\em Russian Math. Surveys}, 52(6):1191--1249, 1997.

\bibitem{MMCP09b}
D.~Maslov, J.~Mathew, D.~Cheung, and D.~K. Pradhan.
\newblock {An $O(m^2)$-depth quantum algorithm for the elliptic curve discrete
  logarithm problem over GF$(2^m)$}.
\newblock {\em Quantum Information \& Computation}, 9(7):610--621, 2009.

\bibitem{MMCP09}
D.~Maslov, J.~Mathew, D.~Cheung, and D.~K. Pradhan.
\newblock {On the Design and Optimization of a Quantum Polynomial-Time Attack
  on Elliptic Curve Cryptography}.
\newblock arXiv:0710.1093v2, February 2009.
\newblock Available at \url{http://arxiv.org/abs/0710.1093v2}.

\bibitem{Mas88}
E.~D. Mastrovito.
\newblock {VLSI designs for multiplication over finite fields $GF(2^m)$}.
\newblock In Teo Mora, editor, {\em Proceedings of the Sixth Symposium on
  Applied Algebra, Algebraic Algorithms and Error Correcting Codes}, volume 357
  of {\em Lecture Notes in Computer Science}, pages 297--309. Springer, 1988.

\bibitem{Mas91}
E.~D. Mastrovito.
\newblock {\em {VLSI Architectures for Computation in Galois Fields}}.
\newblock PhD thesis, Link{\"o}ping University, Link{\"o}ping, Sweden, 1991.

\bibitem{MRR+:2007}
C.~Moore, D.~Rockmore, A.~Russell, and L.~J. Schulman.
\newblock {The power of strong Fourier Sampling: Quantum Algorithms for Affine
  Groups and Hidden Shifts}.
\newblock {\em SIAM Journal on Computing}, 37(3):938--958, 2007.

\bibitem{FIPS1863}
National Institute of Standards and Technology, Gaithersburg, MD 20899-8900.
\newblock {\em FIPS PUB 186-3. Federal Information Processing Standard
  Publication. Digital Signature Standard (DSS)}, June 2009.
\newblock Available at
  \url{http://csrc.nist.gov/publications/fips/fips186-3/fips_186-3.pdf}.

\bibitem{NC:2000}
M.~Nielsen and I.~Chuang.
\newblock {\em Quantum Computation and Quantum Information}.
\newblock Cambridge University Press, 2000.

\bibitem{ORR:2012}
M.~Ozols, M.~Roetteler, and J.~Roland.
\newblock Quantum rejection sampling.
\newblock In {\em Proceedings of the 3rd ACM conference on Innovations in
  Theoretical Computer Science (ITCS'12)}, pages 290--308, 2012.

\bibitem{Reichardt:2009}
B.~W. Reichardt.
\newblock Quantum universality by state distillation.
\newblock {\em Quantum Inf. Comput.}, 9:1030--1052, 2009.

\bibitem{MaHa04}
A.~Reyhani-Masoleh and M.~A. Hasan.
\newblock {Low Complexity Bit Parallel Architectures for Polynomial Basis
  Multiplication over $GF(2^m)$}.
\newblock {\em IEEE Transactions on Computers}, 53(8):945--959, 2004.

\bibitem{RHSCC05}
F.~Rodr{\'{\i}}guez-Henr{\'{\i}}quez, N.~A. Saqib, and N.~Cruz-Cort{\'e}s.
\newblock {A Fast Implementation of Multiplicative Inversion over GF$(2^m)$}.
\newblock In {\em International Symposium on Information Technology: Coding and
  Computing (ITCC 2005)}, volume~1, pages 574--579. IEEE Computer Society,
  2005.

\bibitem{Roetteler:2010}
M.~R{\"o}tteler.
\newblock {Quantum algorithms for highly non-linear {B}oolean functions}.
\newblock In {\em Proceedings of the 21st Annual ACM-SIAM Symposium on Discrete
  Algorithms (SODA'10)}, pages 448--457, 2010.
\newblock Available at \url{http://arxiv.org/abs/0811.3208v2}.

\bibitem{Sho97}
P.~W. Shor.
\newblock {Polynomial-Time Algorithms for Prime Factorization and Discrete
  Logarithms on a Quantum Computer}.
\newblock {\em SIAM Journal on Computing}, 26(5):1484--1509, 1997.

\bibitem{Sil99}
J.~H. Silverman.
\newblock {Fast Multiplication in Finite Fields GF$(2^N)$}.
\newblock In \c{C}etin Kaya~Ko\c{c} and Christof Paar, editors, {\em
  Cryptographic Hardware and Embedded Systems, First International Workshop,
  CHES '99}, volume 1717 of {\em Lecture Notes in Computer Science}, pages
  122--134. Springer, 1999.

\bibitem{TaTa01}
N.~Takagi, J.~Yoshiki, and K.~Takagi.
\newblock {A Fast Algorithm for Multiplicative Inversion in $GF(2^m)$ Using
  Normal Basis}.
\newblock {\em IEEE Transactions on Computers}, 50(5):394--398, 2001.

\end{thebibliography}
\end{document}